\documentclass[12pt]{article}
\usepackage{amsmath}
\usepackage{graphicx}

\def\papertitlepage{\baselineskip 3.5ex\thispagestyle{empty}}
\def\preprinumber#1#2{\hfill\begin{minipage}{4.2cm} #1
        \par\noindent #2 \end{minipage}}
\allowdisplaybreaks[3]

\begin{document}

\papertitlepage
\setcounter{page}{0}
\preprinumber{KEK-TH-1933}{KUNS-2641}
\baselineskip 0.8cm
\vspace*{6ex}

\begin{center}
{\Large\bf Effective Lagrangian in de Sitter Spacetime}
\end{center}

\begin{center}
Hiroyuki K{\sc itamoto}$^{1)}$
\footnote{E-mail address: kitamoto@tap.scphys.kyoto-u.ac.jp}
and
Yoshihisa K{\sc itazawa}$^{2),3)}$
\footnote{E-mail address: kitazawa@post.kek.jp}\\
\vspace{5mm}
$^{1)}$
{\it Division of Physics and Astronomy}\\
{\it Graduate School of Science, Kyoto University}\\
{\it Kyoto 606-8502, Japan}\\
$^{2)}$
{\it KEK Theory Center, Tsukuba, Ibaraki 305-0801, Japan}\\
$^{3)}$
{\it Department of Particle and Nuclear Physics}\\
{\it The Graduate University for Advanced Studies (Sokendai)}\\
{\it Tsukuba, Ibaraki 305-0801, Japan}\\
\end{center}

\vskip 3ex
\baselineskip = 2.5 ex

\begin{center}
{\bf Abstract}
\end{center}

Scale invariant fluctuations of metric are universal feature of quantum gravity in de Sitter spacetime.
We construct an effective Lagrangian which summarizes their implications on local physics by integrating super-horizon metric fluctuations.
It shows infrared quantum effects are local and render fundamental couplings time dependent. 
We impose Lorenz invariance on the effective Lagrangian as it is required by the principle of general covariance. 
We show that such a requirement leads to unique physical predictions by fixing the quantization ambiguities. 
We explain how the gauge parameter dependence of observables is canceled. 
In particular the relative evolution speed of the couplings are shown to be gauge invariant. 



\vspace*{\fill}
\noindent
January 2016

\newpage

\section{Introduction}
\setcounter{equation}{0}

In our quest for the unifying understanding of microscopic and macroscopic physics, we are puzzled by the existence of the large universe itself.
Let us denote the Planck scale by $M_P$, the Hubble constant by $H$, and the characteristic scale of Dark Energy $\Lambda$: 
\begin{align}
H^2 M_P^2 &\sim \Lambda^4,\hspace{1em}\Lambda \sim m_{\nu}, \notag\\
\big(\frac{\Lambda}{M_P}\big)^4&=\big(\frac{H}{M_P}\big)^2\sim 10^{-120}. 
\end{align}
Why is there such a large hierarchy? 
Does Dark Energy change with time? 
How does inflation theory fit in this hierarchy? 
Apparently they are incomprehensible in our present knowledge of field theory and string theory. 
Quantum gravity in de Sitter spacetime is relevant to these questions. 
There still may be large unknown territory in nonperturbative and nonequilibrium physics. 

In de Sitter spacetime, scale independent metric fluctuations are universally generated. 
The two-point function of the tensor mode of the metric $\delta g^T$ exhibits secular growth as more super-horizon modes accumulate with cosmic time $t$: 
\begin{align}
\langle\delta g^T \delta g^T\rangle\simeq \frac{H^2}{M_P^2}\int_{p_\text{min}/a(t)}^H \frac{dP}{P} = \frac{H^2}{M_P^2}\log \big(a(t)/a_0\big) = \frac{H^2}{M_P^2}H(t-t_0), 
\label{RW}\end{align}
where an initial time is introduced as $a_0=e^{Ht_0}=p_\text{min}/H$. 
The metric is jittered randomly by the horizon crossing modes like a dust particle  is jittered by molecules.
It performs Brownian motion with fractal dimension two. 
This IR quantum effect breaks de Sitter symmetry as the correlators are not invariant under a constant translation of cosmic time.
Here we have fixed the minimum comoving momentum cut-off $p_\text{min}$.
In such a case, the physical minimum momentum scales as $P_\text{min}(t)=p_\text{min}/a(t)\propto 1/a(t)$. 
Physically speaking, we consider a situation that a universe with a finite spatial extension starts de Sitter expansion at an initial time $t_0$. 

The secular growth was found in the propagator for a massless and minimally coupled scalar field \cite{VF1982,L1982,S1982}. 
Scale invariant fluctuations are common between the tensor mode of gravity and the light scalar. 
Although we work with the Poincar\'e coordinate, we believe it does not miss important degrees of freedom as the large $t$ behavior (\ref{RW}) also holds in the global coordinate \cite{AF}. 

A possible way to eliminate this secular growth is to fix the physical momentum cut-off $P_\text{min}$.
This procedure may not be consistent with unitarity in a more generic context. 
In inflation theory super-horizon modes may eventually come back into the cosmic horizon. 
It is argued that de Sitter symmetry breaking effects are not observable as the super-horizon modes are pure gauge \cite{HMM,Urakawa2009}. 
A caveat of this argument is that the relevant gauge transformation diverges at spatial infinity, namely it is a large gauge transformation.
We do not throw away super-horizon modes in inflation theory. 
We believe it is a consistent strategy to investigate the entire evolution of a universe of a finite spatial extension when it started de Sitter expansion. 

We have investigated IR logarithmic effects due to the quantum fluctuation of the metric in Schwinger--Keldysh formalism where the both metric and matter are quantized. 
We evaluated the quantum equations of motion, namely the first derivative of the effective action.
The matter fields do not give rise to such effects unless it is a minimally coupled massless scalar field. 
It thus appears to be enough to integrate metric fluctuations only leaving matter fields as classical\footnote{The IR logarithmic effects of a minimally coupled massless scalar field can be included separately and they are manifestly Lorentz invariant.}. 
In the case of the kinetic term, we have considered two types of diagrams: Type A and Type B. 
Type A is the quantum average over the metric fluctuation in the  local vertex. 
Type B corresponds to the intermediate states of the matter and the soft graviton. 
Physically it is difficult to distinguish a particle state from the same state with a soft metric fluctuation. 
In this sense Type B diagram is hard to distinguish from a one particle reducible diagram. 

This fact may be reflected in the quantization ambiguity of the matter field \cite{KitamotoPa}. 
As it turned out, the IR effects depend on the parametrization of the matter field with respect to the conformal metric fluctuation. 
Let us consider a scalar field $\phi$ for example. 
We could rescale the field by the conformal mode of the metric $\omega$ as $e^{\alpha\omega}\phi$. 
The problem is that IR effects are found to be sensitive to this parameter $\alpha$. 
Physically speaking, the matter field is accompanied by soft metric fluctuations and IR effects depend on how to specify them. 
We have proposed that the ambiguity is fixed by requiring that the IR logarithmic effect preserves Lorentz symmetry at a sub-horizon scale. 

We recall here that the propagators of the metric are not Lorentz covariant at the super-horizon scale in de Sitter spacetime. 
We have adopted a non-covariant gauge which provides us the simplest and complete description of the super-horizon modes. 
In this gauge, the conformal mode is not a singlet under full Lorentz transformation but only under spatial rotation. 

Lorentz invariance is one of the fundamental principles of the quantum field theory and microscopic physics. 
In general relativity, Lorentz symmetry must hold at least locally when the spacetime curvature can be ignored. 
In the case of the de Sitter spacetime, it corresponds to the sub-horizon scale. 
In fact the requirement of Lorentz symmetry at this limit follows from the fundamental principle of general relativity. 
From microscopic quantum field theory point of view, general covariance is a necessary consequence of Lorentz invariance for massless spin-2 particles \cite{SW}. 
As the physical degrees of freedom: $\pm 2$ helicity states being the small representation of the Poincar\'{e} group, the general covariance is required to ensure the unitarity of the theory.

We argue that IR logarithmic effects cannot spoil this important symmetry. 
Although it is suppressed by a factor $H^2/M_P^2$, it could become large at late times due to the logarithmic factor $\log a(t)$.
In the literature, different IR logarithmic corrections are reported with different parametrization of the matter field \cite{KitamotoSD,KitamotoG}, \cite{Kahya2007,Miao2005}. 
Nevertheless the unique result is obtained by requiring Lorentz symmetry after reparametrizing the matter field with respect to the conformal mode \cite{KitamotoPa}.

Our requirement of Lorentz invariance at the sub-horizon scale effectively minimizes the IR logarithmic effect. 
In fact we have shown that we can eliminate IR logarithmic effect in the free field theories after time dependent wave function renormalization. 
Nevertheless it is not possible to do so in the interacting theories as couplings acquire time dependence. 

In this paper, we construct an effective Lagrangian in de Sitter spacetime by integrating super-horizon mode of the metric fluctuation. 
In this approach, we consider the contribution from Type A diagram only. 
The Type B diagram contains both local and non-local contributions. 

We postulate that the effect of Type B diagram can be reproduced by a local redefinition of fields as far as IR effects are concerned. 
We represent this effect by rescaling the matter field by the conformal mode $\phi\rightarrow e^{\alpha\omega}\phi$. 
We require that the IR logarithmic effects respect the Lorentz symmetry at the sub-horizon scale. 
This requirement uniquely fixes the rescaling freedom of the matter field. 
We show that this strategy leads to the identical physical predictions with the Schwinger--Keldysh approach under the same requirement of Lorentz symmetry at the sub-horizon scale. 
In this approach, locality of IR logarithms is manifest. 
The validity of our fundamental postulate is underscored by the consistency with the existing results. 

In section 2, we show IR logarithmic effects are local by investing Schwinger--Dyson equations. 
In section 3, we construct an effective Lagrangian. 
In section 4, we examine gauge dependence of physical observables. 
We conclude in section 5. 
We point out a similarity of our strategy to construct an effective Lagrangian in de Sitter spacetime to 2d quantum gravity \cite{DDK}. 
In 2d gravity, there exits a conformal mode dressing ambiguity with respect to an operator $\int d^2x\ e^{2\omega}O(x)\rightarrow \int d^2x\ e^{\beta\omega}O(x)$. 
The dressing parameter $\beta$ is fixed by requiring that this operator becomes conformally invariant. 
This requirement follows from the consistency with the general covariance. 
Here we require Lorentz invariance of the effective Lagrangian at the sub-horizon scale. 
It also follows from the consistency with the general covariance. 
This condition fixes the rescaling ambiguity of the matter fields by the conformal mode in our construction of the effective action.

\section{Locality of IR logarithms}
\setcounter{equation}{0}

In this section, we show that IR logarithmic effects are local as they cancel in non-local contributions. 
We have investigated the two-point function of the conformally coupled scalar field $G^{-+}(x_1,x_2)=\langle\phi (x_1)\phi(x_2)\rangle$ to examine the one-loop metric fluctuation effect \cite{Kitamoto2013}. 
It obeys the following Schwinger--Dyson equation: 
\begin{align}
&\ G_0^{-1}|_{x_1}G^{-+}(x_1,x_2)-G_0^{-1}|_{x_2}G^{-+}(x_1,x_2) \notag\\
=&\ \int d^4x'\ \Sigma_\text{4-pt}(x_1,x')G^{-+}_0(x',x_2) \notag\\
&-\int d^4x'\ G^{-+}_0(x_1,x')\Sigma_\text{4-pt}(x',x_2) \notag\\
&+\int d^4x'\ \Sigma^R_\text{3-pt}(x_1,x')G^{-+}_0(x',x_2)+\int d^4x'\ \Sigma^{-+}_\text{3-pt}(x_1,x')G^A_0(x',x_2) \notag\\
&-\int d^4x'\ G^{-+}_0(x_1,x')\Sigma^A_\text{3-pt}(x',x_2)-\int d^4x'\ G^R_0(x_1,x')\Sigma^{-+}_\text{3-pt}(x',x_2),  
\label{B0}\end{align}
where $G_0^{-1}=i(\partial_0^2-\partial_i^2)$ is the Laplacian and $\Sigma$ denotes the self-energy of the scalar field due to the metric fluctuation. 
$R$ and $A$ denote retarded and advanced Green functions respectively.

In momentum space, 
\begin{align}
&2i\partial_{\tau_c}\partial_{\Delta \tau}G^{-+}(\tau_1,\tau_2,\vec{p}) \notag\\
=&\ \Sigma_\text{4-pt}(\tau_1)G^{-+}_0(\tau_1,\tau_2,\vec{p}) 
- G^{-+}_0(\tau_1,\tau_2,\vec{p})\Sigma_\text{4-pt}(\tau_2) \notag\\
&+\int d\tau'\ \Sigma^R_\text{3-pt}(\tau_1,\tau',\vec{p})G^{-+}_0(\tau',\tau_2,\vec{p})
+\int d\tau'\ \Sigma^{-+}_\text{3-pt}(\tau_1,\tau',\vec{p})G^A_0(\tau',\tau_2,\vec{p}) \notag\\
&-\int d\tau'\ G^R_0(\tau_1,\tau',\vec{p})\Sigma^{-+}_\text{3-pt}(\tau',\tau_2,\vec{p})
-\int d\tau'\ G^{-+}_0(\tau_1,\tau',\vec{p})\Sigma^A_\text{3-pt}(\tau',\tau_2,\vec{p}). 
\label{Bmn}\end{align}
where $\tau_c=(\tau_1+\tau_2)/2$ and $\Delta \tau=\tau_1-\tau_2$.
This is a standard procedure to derive Boltzmann type equations. 
The left-hand side picks up $\tau_c$ dependence of the propagator. 
It is caused by the ``collision terms'' on the right-hand side. 

They are integrated over the conformal time of the interaction vertices.
The self-energy part $\Sigma$ contains the IR singularity $\int d^3q/q^3$ due to super-horizon mode of the metric. 
We focus on the non-local contributions in the ``collision terms''. 
The IR singularity of them can be estimated as\footnote{In our gauge, IR singularity in $\Sigma_\text{non-local}$ cancels as  $\langle h^{\mu\nu}h^{\rho\sigma}(q)\rangle p_\mu p_\nu p_\rho p_\sigma \sim 0$.  
It is no longer the case in a different gauge.} 
\begin{align}
\Sigma_\text{non-local}\sim \int d^3q\ \langle h^{\mu\nu}h^{\rho\sigma}(q)\rangle p_{\mu}p_{\nu}p_{\rho}p_{\sigma}G_0(\tau_1,\tau_2,\vec{p}). 
\label{NLSE}\end{align}

First we observe that such IR singularities cancel in the last two lines of (\ref{Bmn}) 
\begin{align}
&\int d^3q\ \langle h^{\mu\nu}h^{\rho\sigma}(q)\rangle p_{\mu}p_{\nu}p_{\rho}p_{\sigma}\times \notag\\
&\Big\{\int d\tau'\ G_0^R(\tau_1,\tau',\vec{p})G^{-+}_0(\tau',\tau_2,\vec{p})
+\int d\tau'\ G^{-+}_0(\tau_1,\tau',\vec{p})G^A_0(\tau',\tau_2,\vec{p}) \notag\\
&-\int d\tau'\ G^R_0(\tau_1,\tau',\vec{p})G^{-+}_0(\tau',\tau_2,\vec{p})
-\int d\tau'\ G^{-+}_0(\tau_1,\tau',\vec{p})G^A_0(\tau',\tau_2,\vec{p})\Big\}. 
\end{align}
They represent a particle $\rightarrow$ a particle $+$ a soft graviton and the inverse process. 
The signs are opposite since the process and the inverse process decrease and increase the weight of a particle state. 
The cancellation takes place as we cannot distinguish a particle and a particle with a soft graviton.

Second, the integration over the conformal time could still produce IR divergences from large negative region.
They also cancel between real and virtual contributions since such a cancellation can be manifestly demonstrated in the last two lines of (\ref{Bmn}) as
\begin{align}
\int_{\tau_2}^{\tau_1} d\tau'\ \Sigma^{-+}_\text{3-pt}(\tau_1,\tau',\vec{p})G^{-+}_0(\tau',\tau_2,\vec{p})
-\int_{\tau_2}^{\tau_1}  d\tau'\ G^{-+}_0(\tau_1,\tau',\vec{p})\Sigma^{-+}_\text{3-pt}(\tau',\tau_2,\vec{p}). 
\label{RVC}\end{align}
On the other hand it has been explicitly shown that the local contributions lead to IR logarithmic effects: 
\begin{align}
&\partial_{\tau_c}Z=\frac{3}{8}\frac{\kappa^2H^2}{4\pi^2}\frac{1}{\tau_c}, \notag\\
&Z=1-\frac{3}{8}\frac{\kappa^2H^2}{4\pi^2}\log a(\tau_c), 
\label{Z}\end{align}
where $Z$ is the overall normalization (wave function renormalization) factor of the propagator. 

Let us consider the perturbation by a constant metric $\bar{h}^{\mu\nu}$. 
Namely we replace the super-horizon mode by a constant metric. 
The cubic term is
\begin{align}
\frac{1}{2}\int d^4x\ \bar{h}^{\mu\nu}\partial_{\mu}\phi\partial_{\nu}\phi. 
\end{align}
The second order self-energy is
\begin{align}
\Sigma \sim \bar{h}^{\mu\nu}\bar{h}^{\rho\sigma}p_{\mu}p_{\nu}p_{\rho}p_{\sigma}G_0(\tau_1,\tau_2,\vec{p}). 
\end{align}
The time dependence of the two-point function is estimated as
\begin{align}
&\bar{h}^{\mu\nu}\bar{h}^{\rho\sigma}p_{\mu}p_{\nu}p_{\rho}p_{\sigma}\times \notag\\
&\Big\{\int d\tau'\ G_0^R(\tau_1,\tau',\vec{p})G^{-+}_0(\tau',\tau_2,\vec{p})
+\int d\tau'\ G^{-+}_0(\tau_1,\tau',\vec{p})G^A_0(\tau',\tau_2,\vec{p}) \notag\\
&-\int d\tau'\ G^R_0(\tau_1,\tau',\vec{p})G^{-+}_0(\tau',\tau_2,\vec{p})
-\int d\tau'\ G^{-+}_0(\tau_1,\tau',\vec{p})G^A_0(\tau',\tau_2,\vec{p})\Big\}. 
\end{align}
Therefore the cancellation of IR logarithm: $\log a(\tau_c)$ in non-local contributions holds if $\tau_c$ dependence cancels with a constant metric perturbation. 
The latter cancellation follows from the time translation invariance with a constant metric perturbation. 
It is clear that this argument can be generalized beyond the one-loop level.

We can make a correspondence between non-local contributions of the IR metric fluctuations and 
those of the constant metric perturbation to all orders through identifying super-horizon modes with a constant background.
So the cancellation of the IR singularity in the non-local contributions follows from the absence of $\tau_c$ dependence due to a constant background. 
Hence the former is suppressed by a factor $\Delta \tau/\tau_c\ll 1$ in the sub-horizon scale where the IR logarithmic factor  $\log a$ changes slowly. 
It is because super-horizon modes are not quite constant but jolted randomly with the characteristic scale of $H$. 
We conclude that IR logarithms must cancel in the non-local contributions in generic situations.  

In order to understand the cancellation mechanism of IR logarithms in non-local contributions, we may consider a simpler model. 
We replace metric fluctuations by that of the minimally coupled scalar field $\varphi$ with a cubic interaction to $\phi$ 
\begin{align}
\lambda_3\int d^4x\ a^2(\tau ) \varphi\phi^2.
\end{align}
The Schwinger--Dyson equation is
\begin{align}
&2i\partial_{\tau_c}\partial_{\Delta \tau}G^{-+}(\tau_1,\tau_2,\vec{p}) \notag\\
=&\ \int d\tau'\ \Sigma^R(\tau_1,\tau',\vec{p})G^{-+}_0(\tau',\tau_2,\vec{p})
+\int d\tau'\ \Sigma^{-+}(\tau_1,\tau',\vec{p})G^A_0(\tau',\tau_2,\vec{p}) \notag\\
&-\int d\tau'\ G^R_0(\tau_1,\tau',\vec{p})\Sigma^{-+}(\tau',\tau_2,\vec{p})
-\int d\tau'\ G^{-+}_0(\tau_1,\tau',\vec{p})\Sigma^A(\tau',\tau_2,\vec{p}), 
\label{BYK}\end{align}
where $\Sigma$ has the following IR singularity
\begin{align}
\Sigma(\tau_1,\tau_2,\vec{p})\sim \lambda_3^2\int \frac{d^3q}{q^3}\ \frac{1}{\tau_1^2\tau_2^2}G_0(\tau_1,\tau_2,\vec{p}). 
\end{align}
We observe again that the IR singularity in the integrand cancel on the right-hand side of (\ref{BYK}) at sub-horizon scale where $\Delta \tau/\tau_c \ll1$ 
\begin{align}
&\lambda_3^2\int \frac{d^3q}{q^3}\times \notag\\
&\Big\{\int d\tau'\ \frac{1}{\tau_1^2\tau'^2}G^R_0(\tau_1,\tau',\vec{p})G^{-+}_0(\tau',\tau_2,\vec{p})
+\int d\tau'\ \frac{1}{\tau_1^2\tau'^2}\ G^{-+}_0(\tau_1,\tau',\vec{p})G^A_0(\tau',\tau_2,\vec{p}) \notag\\
&-\int d\tau'\ \frac{1}{ \tau'^2\tau_2^2}\ G^R_0(\tau_1,\tau',\vec{p})G^{-+}_0(\tau',\tau_2,\vec{p})
-\int d\tau'\ \frac{1}{\tau'^2\tau_2^2}\ G^{-+}_0(\tau_1,\tau',\vec{p})G^A_0(\tau',\tau_2,\vec{p})\Big\}. 
\end{align}
The integration of the conformal time does not give rise to IR singularity as an analogous formula with (\ref{RVC}) holds.

Note that $\Sigma$ is identical with the second order self-energy due to a mass perturbation under the identification $m^4=\lambda_3^2\int d^3q/q^3$. 
Therefore the cancellation of IR logarithms holds in this model if $\tau_c$ dependence cancels with a mass perturbation. 
The latter cancellation is a consequence of de Sitter symmetry of a mass perturbation. 
de Sitter symmetry implies no time dependence of microscopic physics as $\tau_c$ can be scaled out.
We conclude that the cancellation of IR logarithms with the cubic interaction follows from this correspondence. 

Note that the mass term can be identified with the constant expectation value of the scalar field in the cubic interaction $m^2\sim \lambda_3\bar{\varphi}$. 
We can make a correspondence between diagrams of the cubic interaction and those of the mass perturbation to all orders through this identification 
\begin{align}
m^{2n}\sim  \lambda_3^{n}\bar{\varphi}^{n}. 
\end{align}
So the cancellation of the IR singularity in the non-local
contributions follows from the absence of $\tau_c$ dependence due to the constant background.
Since a constant background preserves de Sitter symmetry, it must be the case in general.

Let us come back to the IR logarithms due to the metric fluctuations. 
It remains the same that the cancellation of the IR singularity in the non-local contributions follows from the absence of $\tau_c$ dependence due to a constant background metric perturbation. 
It also remains true that no $\tau_c$ dependence arise for a constant background metric perturbation since it preserves de Sitter symmetry. 
We conclude that IR logarithms must cancel in the non-local contributions with metric fluctuations in generic situations.

It is not possible to distinguish one particle state from the same state with a soft graviton. 
The situation is similar to one particle state with soft photons in QED \cite{Kinoshita1962,Lee1964}. 
It is the physical reason behind the cancellation of IR logarithms in the non-local contributions. 
The cancellation of non-local IR logarithms follows from the absence of $\tau_c$ dependence with a constant background. 
We believe that the cancellation of IR divergences at large negative conformal time region is a universal phenomena in unitary field theories. 

On the other hand, there is no such cancellation mechanism for local contributions. 
The self-energy for the local contributions can be estimated as 
\begin{align}
\Sigma_\text{local}(x,x')\sim i\delta^{(4)}(x-x')\times\big\{\log \big(a(\tau')/a_0\big) \partial'^2-\partial'_0\log \big(a(\tau')/a_0\big)\cdot\partial'_0\big\}, 
\end{align}
where the IR logarithm comes from the gravitational propagator at the coincident point. 
Since the IR logarithm is not quite constant, the time derivative of it is nonzero. 
Considering partial integrations, we can confirm that the local contributions do not cancel out: 
\begin{align}
&\ \int d^4x'\ \Sigma_\text{local}(x_1,x')G^{-+}(x',x_2)-\int d^4x'\ G^{-+}(x_1,x')\Sigma_\text{local}(x',x_2) \notag\\
\sim&-2iHa(\tau_c)\partial_0|_{x_1}G(x_1,x_2). 
\end{align}
That is how the local contributions lead to the time dependent normalization factor (\ref{Z}). 
The IR logarithmic behavior of the propagator is a consequence of the accumulation of the super-horizon modes with cosmic evolution. 
We therefore argue that the locality of IR logarithmic effect holds in a generic situation. 

\section{Effective Lagrangian}\label{Effective}
\setcounter{equation}{0}

In this section, we construct an effective Lagrangian in de Sitter spacetime which incorporates 
IR logarithmic effects.  
The de Sitter metric is
\begin{align}
ds^2&=-dt^2+a^2(t)dx^idx_i \notag\\
&=a^2(\tau)(-d\tau^2+dx^idx_i). 
\end{align}
We parametrize the metric including the quantum fluctuation as
\begin{align}
g_{\mu\nu}&=a^2(\tau )e^{2\omega}\tilde{g}_{\mu\nu}, \notag\\
\text{det}\ \tilde{g}_{\mu\nu}&=-1,\hspace{1em}\tilde{g}_{\mu\nu}= (e^h)_{\mu\nu}, 
\end{align}
where $h_{\mu\nu}$ and $\omega$ represent traceless and conformal modes of the metric respectively.
In this paper, the Lorentz indices are raised and lowered by the flat metric $\eta^{\mu\nu}$ and $\eta_{\mu\nu}$ respectively. 

The Lagrangian of Einstein gravity on the $4$-dimensional de Sitter background is 
\begin{align}
\mathcal{L}_\text{gravity}&=\frac{1}{\kappa^2}\sqrt{-g}\big[R-6H^2\big] \notag\\
&=\frac{1}{\kappa^2}\big[\Omega^2\tilde{R} 
+6\tilde{g}^{\mu\nu}\partial_\mu\Omega\partial_\nu\Omega-6H^2\Omega^4\big], 
\label{gravity}\end{align}
where $\tilde{R}$ denotes the Ricci scalar constructed from $\tilde{g}_{\mu\nu}$. 
The gravitational coupling $\kappa$ is related to Newton's constant as $\kappa^2=16\pi G_N$.

In order to fix the gauge with respect to general coordinate invariance, we adopt the following gauge fixing term \cite{Tsamis1992}: 
\begin{align}
\mathcal{L}_\text{GF}&=-\frac{1}{2}a^{2}F_\mu F^\mu, \notag\\
F_\mu&=\partial_\rho h_\mu^{\ \rho}-2\partial_\mu \omega+2h_\mu^{\ \rho}\partial_\rho\log a+4\omega\partial_\mu\log a. 
\label{GF}\end{align}

We decompose the spatial part of the metric as
\begin{align}
h^{ij}=\tilde{h}^{ij}+\frac{1}{3}h^{kk}\delta^{ij}=\tilde{h}^{ij}+\frac{1}{3}h^{00}\delta^{ij}. 
\end{align} 
After diagonalizing the quadratic action in terms of
\begin{align}
X=2\sqrt{3}\omega-\frac{1}{\sqrt{3}}h^{00},\hspace{1em}Y=h^{00}-2\omega, 
\label{diagonalize}\end{align}
we find that some metric modes behave as the massless and minimally coupled scalar field $\varphi$ and the other modes behaves as the massless and conformally coupled mode $\phi$:    
\begin{align}
\langle X(x)X(x')\rangle&=-\langle\varphi(x)\varphi(x')\rangle, \notag\\
\langle\tilde{h}^i_{\ j}(x)\tilde{h}^k_{\ l}(x')\rangle&=(\delta^{ik}\delta_{jl}+\delta^i_{\ l}\delta_j^{\ k}-\frac{2}{3}\delta^i_{\ j}\delta^k_{\ l})\langle\varphi(x)\varphi(x')\rangle, \notag\\
\langle b^i(x)\bar{b}^j(x')\rangle&=\delta^{ij}\langle\varphi(x)\varphi(x')\rangle, 
\label{minimally}\end{align}
\begin{align}
\langle h^{0i}(x)h^{0j}(x')\rangle&=-\delta^{ij}\langle\phi(x)\phi(x')\rangle, \notag\\
\langle Y(x)Y(x')\rangle&=\langle\phi(x)\phi(x')\rangle, \notag\\
\langle b^0(x)\bar{b}^0(x')\rangle&=-\langle\phi(x)\phi(x')\rangle, 
\label{conformally}\end{align}
where $b$, $\bar{b}$ denote the ghost and anti-ghost fields. 

Since we focus on the de Sitter symmetry breaking effects, we may introduce an approximation. 
We can neglect the conformally coupled modes of gravity (\ref{conformally}) since they do not induce the IR logarithm. 
In such an approximation, the following identity holds 
\begin{align}
h^{00}\simeq 2\omega\simeq\frac{\sqrt{3}}{2}X. 
\label{diagonalize1}\end{align}
Note that we are still left with scalar $X$ and spin-2 modes $\tilde{h}^{ij}$ which contain the tensor modes.
Their total contribution is found to preserve Lorentz symmetry at sub-horizon scale in this gauge.

We consider a generic renormalizable Lagrangian just like the standard model: 
\begin{align}
\int \sqrt{-g}d^4x\ \big[&-g^{\mu\nu}D_{\mu}\phi (D_{\nu}\phi)^*-(\frac{1}{6}R+m^2)\phi\phi^*+i\bar{\psi}e^\mu_{\ a}\gamma^aD_{\mu}\psi - m_f\bar{\psi}\psi \notag\\
&-\frac{1}{2}\lambda_4(\phi\phi^*)^2 -\lambda_Y\phi\bar{\psi}\psi+\text{(h.c.)}-\frac{1}{4g^2}g^{\mu\rho}g^{\nu\sigma}F^a_{\mu\nu}F^a_{\rho\sigma}\big]. 
\end{align}
We assume that there are no massless minimally coupled scalar field.

We propose to construct the effective Lagrangian as
\begin{align}
\mathcal{L}_\text{eff}=\langle\mathcal{L}\rangle_\text{metric}, 
\label{EL}\end{align}
where the average is taken over the super-horizon metric fluctuation only.
The effective Lagrangian $\mathcal{L}_\text{eff}$ summarizes IR logarithmic effects in de Sitter spacetime. 
The justification of this proposal is that IR logarithmic effects are contained in the metric fluctuations only. 
The effect of the matter field fluctuation should be neglected except possible local redefinition of fields. 
When we construct the effective action, we rescale the fields as
\begin{align}
\phi &\rightarrow a^{-1}e^{(\alpha-1)\omega}\phi, \notag\\
\psi &\rightarrow a^{-\frac{3}{2}}e^{(\beta-\frac{3}{2})\omega}\psi, \notag\\
A_{\mu} &\rightarrow A_{\mu}. 
\label{rescale}\end{align}
We point out that the gauge field is protected against the rescaling ambiguity due to gauge symmetry.  

After the field redefinition, we obtain
\begin{align}
\int d^4x\ \big[&-e^{2\alpha\omega}\tilde{g}^{\mu\nu}D_{\mu}\phi (D_{\nu}\phi)^*-m^2a^2e^{2(1+\alpha)\omega}\phi\phi^* 
+ie^{2\beta\omega}\bar{\psi}\tilde{e}^\mu_{\ a}\gamma^aD_{\mu}\psi + m_fae^{(1+2\beta)\omega}\bar{\psi}\psi \notag\\
&-\frac{1}{2}e^{4\alpha\omega}\lambda_4(\phi\phi^*)^2 -\lambda_Ye^{(\alpha+2\beta)\omega}\phi\bar{\psi}\psi+\text{(h.c.)}
-\frac{1}{4g^2}\tilde{g}^{\mu\rho}\tilde{g}^{\nu\sigma}F^a_{\mu\nu}F^a_{\rho\sigma}\big]. 
\label{rescaled}\end{align}
We have neglected derivatives of the conformal mode as we focus on the IR effects.
Our strategy is to adjust $\alpha$ and $\beta$ to preserve Lorentz invariance at the sub-horizon scale.
The metric propagators are suppressed by $H^2/M_P^2$ factors.
So the leading IR effects appear as $(H^2/M_P^2\cdot\log a)^n$ at the $n$-th loop.
We investigate the leading one-loop effect. 

As an illustration, let us consider the kinetic term of the scalar field in (\ref{rescaled}) 
\begin{align}
&-\langle e^{2\alpha\omega}\tilde{g}^{\mu\nu}\rangle D_{\mu}\phi (D_{\nu}\phi)^* \notag\\
\sim &-\langle e^{2\alpha\omega}\tilde{g}^{00}\rangle D_0\phi (D_0\phi)^* 
-\langle e^{2\alpha\omega}\tilde{g}^{ij}\rangle D_i\phi (D_j\phi)^*. 
\end{align}
The $\alpha=0$ contribution is 
\begin{align}
-\frac{3}{8}\langle\varphi^2\rangle D_0\phi (D_0\phi)^*-\frac{13}{8}\langle\varphi^2\rangle D_i\phi (D^i\phi)^*, 
\end{align}
where $\langle\varphi^2\rangle=\frac{\kappa^2H^2}{4\pi ^2}\log a(\tau )$. 
The linear term in $\alpha$ is 
\begin{align}
-\frac{3\alpha}{4}\langle\varphi^2\rangle D_0\phi (D_0\phi)^*-\frac{\alpha}{4}\langle \varphi^2\rangle D_i\phi (D^i\phi)^*. 
\end{align}
We find that the requirement of Lorentz invariance fixes $\alpha=-2$.
The total result including $\alpha^2$ effect is
\begin{align}
-\frac{3}{8}\langle\varphi^2\rangle D_0\phi (D_0\phi)^*+\frac{3}{8}\langle\varphi^2\rangle D_i\phi (D^i\phi)^*. 
\end{align}
We thus find that the IR logarithmic effect can be canceled by the time dependent wave function renormalization of 
$\phi\rightarrow Z\phi$ where $Z^2=(1+\frac{3}{8}\langle\varphi^2\rangle)$.

As we have seen, the IR logarithmic effects can be eliminated for free fields in this way. 
However it is not so for interacting fields. 
Let us consider the scalar quartic coupling in our parametrization with canonically normalized kinetic term 
\begin{align}
\lambda_4 (\phi\phi^*)^2 \rightarrow \lambda_4 Z^4e^{-8\omega}(\phi\phi^*)^2. 
\end{align}
We find that the coupling decreases with time
\begin{align}
\lambda_4 Z^4\langle e^{-8\omega}\rangle=\lambda_4(1-\frac{21}{4}\langle\varphi^2\rangle), 
\end{align}
in agreement with \cite{KitamotoSD}.

We next consider the mass term
\begin{align}
m^2a^2e^{2\omega}\phi\phi^* \rightarrow m^2a^2 e^{-2\omega}Z^2\phi\phi^*. 
\end{align}
Since $\langle e^{-2\omega}\rangle Z^2\sim 1$, the mass term is not renormalized after the wave function renormalization. 

Subsequently we consider the kinetic term of the Dirac field in (\ref{rescaled})
\begin{align}
S=\int d^4x\ ie^{2\beta\omega}\bar{\psi}\tilde{e}^\mu_{\ a}\gamma^a D_\mu\psi. 
\label{Dirac}\end{align}
The analogous considerations with the the scalar field fixes $\beta=-1$. 
The IR logarithmic contribution to the kinetic term is 
\begin{align}
-\frac{3}{32}\langle \varphi^2\rangle i\bar{\psi}\gamma^{\mu} D_\mu\psi. 
\end{align}
This effect can be canceled after the time dependent wave function renormalization of $\psi\rightarrow Z_{\psi}\psi$ as $Z_{\psi}^2=(1+\frac{3}{32}\langle\varphi^2\rangle)$. 

We consider the Yukawa coupling in our parametrization with the canonically normalized kinetic terms for the both scalar and Dirac fields 
\begin{align}
\lambda_Y \phi\bar{\psi}\psi\rightarrow \lambda_Y Z Z_\psi^2e^{-4\omega}\phi\bar{\psi}\psi. 
\end{align}
We find that the Yukawa coupling decreases with time
\begin{align}
\lambda_Y Z Z_\psi^2\langle e^{-4\omega}\rangle=\lambda_Y (1-\frac{39}{32}\langle\varphi^2\rangle), 
\end{align}
in agreement with \cite{KitamotoSD}. 

We next consider the mass term
\begin{align}
m_f ae^{\omega}\bar{\psi}\psi \rightarrow m_f ae^{-\omega}Z_{\psi}^2\bar{\psi}\psi. 
\end{align}
Since $\langle e^{-\omega}\rangle Z_{\psi}^2\sim 1$, the mass term is not renormalized after the wave function renormalization.

These considerations parallel our physical interpretations of the IR logarithmic effects in terms of the conformal mode dynamics \cite{review}. 
This work may be regarded as a more precise statement of such an idea. 

We consider the kinetic term of the gauge field in (\ref{rescaled}) 
\begin{align}
-\frac{1}{4g^2}\langle\tilde{g}^{\mu\rho}\tilde{g}^{\nu\sigma}\rangle F^a_{\mu\nu}F^a_{\rho\sigma}=-\frac{1}{4g^2}(1+\frac{3\kappa H^2}{8\pi^2}\log a(\tau))F^a_{\mu\nu}F^{a\mu\nu}. 
\end{align}
Thus IR effects screen the gauge coupling with time.
\begin{align}
g^2(\tau) =g^2(1-\frac{3\kappa H^2}{8\pi^2}\log a(\tau)). 
\end{align}
The one-loop IR effect is to make the gauge coupling time dependent
\begin{align}
-\frac{1}{4g^2(\tau )}F^a_{\mu\nu}F^{a\mu\nu}-D_{\mu}\phi (D^{\mu}\phi)^*+i\bar{\psi }\gamma^{\mu}D_{\mu}\psi. 
\end{align}
It is because the gauge field appears just like the derivatives in the covariant derivatives. 
It shows the consistency of our result with the gauge invariance. 
This conclusion is in agreement with \cite{KitamotoG} given in a background gauge. 
It clearly shows that the IR logarithmic effect is gauge invariant and local. 

As we have shown, the effective Lagrangian approach universally applies to scalar, spinor and vector fields. 
The advantage is its simplicity and robustness of the conclusion. 
It clearly shows that IR logarithmic effect is local. 
The requirement of Lorentz invariance at sub-horizon scale minimizes the IR logarithmic effects in such a way that it disappears in the free field theories. 
Nevertheless it makes couplings to evolve with time  in interacting theories. 

\section{Gauge dependence}
\setcounter{equation}{0}

So far, we have worked in a particular gauge.
Although the two-point functions of the tensor mode are gauge invariant, it is not the case for the other modes of the metric. 
Nevertheless the contributions from tensor mode alone cannot lead to the Lorentz invariant result. 
In other words, Lorentz invariance requires contributions from both the gauge invariant and gauge dependent degrees of freedom. 
So there is no Lorentz invariant quantity per se which does not depend on the gauge parameter. 
In order to show that the time evolution of the couplings is physical, we need to identify relations among them in which the gauge dependence cancels out. 

In order to investigate this question, we consider a more general gauge with a parameter $\beta$: 
\begin{align}
\mathcal{L}_\text{GF}&=-\frac{1}{2}a^{2}F^{\beta}_\mu F^{\beta\mu}, \notag\\
F^{\beta}_\mu&=\beta(\partial_\rho h_\mu^{\ \rho}-2\partial_\mu \omega)+\frac{1}{\beta} (2h_\mu^{\ \rho}\partial_\rho\log a+4\omega\partial_\mu\log a). 
\label{generalGF}\end{align}

As it turns out, the evolution speed of each coupling depends on a gauge parameter $\beta$. 
However the gauge dependence cancels out in the ratio of the evolution speed of the couplings at least for an infinitesimal change of the gauge parameter  $\beta \sim 1$. 
To the first order of $\delta =\beta^2-1$,
\begin{align}
&\langle h^{00}h^{00}\rangle = (1-\delta)\langle h^{00}h^{00}\rangle_0, \notag\\
&\langle\tilde{h}^{ij}\tilde{h}^{kl}\rangle=(1-\delta)\langle\tilde{h}^{ij}\tilde{h}^{kl}\rangle_0, 
\label{P-Scaling}\end{align}
where $\langle h^{00}h^{00}\rangle_0$ denotes the correlator in the original gauge $\delta=0$. 
To evaluate the effective Lagrangian (\ref{EL}), these propagators are investigated at a coincident point. 

From (\ref{P-Scaling}), the time evolutions of the effective couplings are evaluated as   
\begin{align}
\lambda_4(\tau)/\lambda_4 = f^{\Delta_4}(\tau),\hspace{1em}\lambda_Y(\tau)/\lambda_Y=f^{\Delta_Y}(\tau),\hspace{1em}g(\tau)/g=f^{\Delta_g}(\tau), 
\label{Scaling}\end{align}
\begin{align}
\Delta_4=\frac{21}{4},\hspace{1em}\Delta_Y=\frac{39}{32},\hspace{1em}\Delta_g=\frac{3}{4}, 
\label{Exponents}\end{align}
where $f(\tau)$ is a gauge dependent decaying function: 
\begin{align}
f(\tau)=1-(1-\delta)\frac{\kappa^2 H^2}{4\pi^2}\log a(\tau). 
\end{align} 
It is no surprise that the time dependence of each effective coupling is gauge dependent. 
That is because time is observer-dependent as we may reparametrize it.
Our proposal is that the ratio of the evolution speed of the couplings are the observables in de Sitter spacetime \cite{KitamotoSD,KitamotoG}. 
Specifically, we can construct gauge invariant quantities as\footnote{The scaling exponents $\Delta_i$ themselves are gauge dependent. 
We can set $f(\tau)$ as $f(\tau)=1-\frac{\kappa^2 H^2}{4\pi^2}\log a(\tau)$, and then they become $\delta$-dependent.} 
\begin{align}
\Delta_Y/\Delta_4=\frac{13}{56},\hspace{1em}\Delta_g/\Delta_4=\frac{1}{7}. 
\label{Ratios}\end{align}
The gauge invariance of the relative scaling exponents can be interpreted as follows: 
It is sensible to pick a particular coupling and use its time evolution as a physical time. 
We assigned the role to the coupling of the quartic interaction in (\ref{Ratios}). 
In this setting, the relative scaling exponents measure the time evolution of the couplings in terms of a physical time. 

Such a relation is reminiscent of the following phenomena in non-equilibrium physics.
In quantum quench when a system is suddenly brought into criticality,
the one-point functions behave as \cite{Calabrese}
\begin{align}
\langle O_i\rangle=\tilde{f}^{\tilde{\Delta}_i}(\tau), 
\end{align}
where $\tilde{\Delta}_i$ is the conformal dimension of $O_i$ and $\tilde{f}(\tau)$ is a non-universal decaying function. 
The ratios of relaxation times of different operators are universal in $1+1$ dimension. 

It is important to investigate whether this idea works to all orders of gauge parameter shift $\delta$.
To the second order of $\delta$, we find in the Appendix
\begin{align}
&\langle h^{00}h^{00}\rangle_{\delta} = (1-\delta-\frac{5}{2}\delta^2)\langle h^{00}h^{00}\rangle_0, \notag\\
&\langle\tilde{h}^{ij}\tilde{h}^{kl}\rangle_{\delta}=(1-\delta+\frac{5}{4}\delta^2)\langle\tilde{h}^{ij}\tilde{h}^{kl}\rangle_0. 
\end{align}
The gauge parameter no longer cancels in the ratio of two independent correlators.
We need a further consideration to circumvent this problem.

Under this circumstance, we may reparametrize the matter field such that
\begin{align}
\phi \rightarrow e^{\gamma\tau\partial_0}\phi,\hspace{1em}
\psi \rightarrow e^{\gamma\tau\partial_0}\psi,\hspace{1em}
A_0^a \rightarrow e^\gamma e^{\gamma\tau\partial_0}A_0^a,\hspace{1em} 
A_i^a \rightarrow e^{\gamma\tau\partial_0}A_i^a. 
\end{align}
This may be interpreted as a reparametrization of the conformal time: 
\begin{align}
\tau \rightarrow e^{\gamma} \tau. 
\label{fieldrepara}\end{align}
Using this freedom, we may modify the way $h^{00}$, $\omega$ couple to the matter fields as\footnote{This kind of counter term is necessary to renormalize UV divergences in a non-covariant gauge \cite{Kitamoto2013}.} 
\begin{align}
h^{00} \rightarrow h^{00} +\frac{3}{2}\gamma,
\hspace{1em} \omega \rightarrow \omega+\frac{3}{4}\gamma. 
\label{metrictrans}
\end{align}
Since $\partial_{\mu}(e^\gamma \tau )=e^\gamma\delta_\mu^{\ 0}+O(k\tau)$, it is a small transformation for super-horizon mode. 
The same cannot be said for a reparametrization of spatial coordinates $x^i\rightarrow (e^h)^i_{\ j} x^j$. 
Here $\partial_\mu((e^h)^i_{\ j} x^j)=(e^h)^i_{\ j}\delta_\mu^{\ j}+O(k x^i)$, the $O(kx^i)$ term cannot be neglected. 
Thus we cannot transform tensor modes in an analogous way. 

We can choose $\gamma=\frac{5}{4}\delta^2h^{00}$ in such a way that the gauge dependence cancels in the ration of two independent correlators: $h^{00}_\text{eff}=2\omega_\text{eff}=(1+\frac{15}{8}\delta^2)h^{00}$ and $\tilde{h}^{ij}$. 
We also need to modify the rescaling of the fields as 
\begin{align}
\phi &\rightarrow a^{-1}e^{(\alpha-1)\omega_\text{eff}}\phi, \notag\\
\psi &\rightarrow a^{-\frac{3}{2}}e^{(\beta-\frac{3}{2})\omega_\text{eff}}\psi,  \notag\\
A_{\mu} &\rightarrow A_{\mu}. 
\label{rescale'}\end{align}
As far as the IR logarithmic effects to the matter fields are concerned, the Lorentz invariance is preserved at sub-horizon scale in this procedure.
Furthermore the scaling relation (\ref{Scaling})-(\ref{Exponents}) holds to the second order of gauge parameter change $\delta$ with $f(\tau)=1-(1-\delta+\frac{5}{4}\delta^2)\frac{\kappa^2 H^2}{4\pi^2}\log a(\tau)$. 
Since we have one parameter freedom $\gamma$ for the gauge parameter change $\delta$, we can repeat the same procedure to all orders of $\delta$. 
This argument provides a further evidence that the scaling relation (\ref{Scaling})-(\ref{Exponents}) is observable in de Sitter spacetime. 

The essential point here is that there are two independent correlators of metric: scalar and spin-2 modes. 
The gauge dependence of the relative magnitude can be canceled by a reparametrization of the matter field. 
Such a procedure is required by the Lorentz invariance of matter dynamics at sub-horizon scale. 
So Lorentz invariance leads to gauge independence of the observables. 

Let us consider the graviton propagator in a different class of gauge: namely covariant de Donder gauge \cite{deDondergauge}.
In this gauge the two-point function of spin-2 mode $\tilde{h}^{ij}$ is identical to that in the original gauge with $\delta=0$ 
\begin{align}
\langle\tilde{h}^{ij}\tilde{h}^{kl}\rangle_\text{de Donder}=\langle\tilde{h}^{ij}\tilde{h}^{kl}\rangle_0. 
\end{align}
However the remaining (scalar) component of the propagator in de Donder gauge does not agree with that in the original gauge $\delta=0$. 
In fact it exhibits more singular IR behavior. 
We focus on the traceless part of the propagator as it is relevant to possible Lorentz symmetry breaking.
The trace part decouples when the matter field is conformally coupled.  
The important point is that the IR singular part of this component is pure gauge. 
We can effectively eliminate that part of the graviton propagator by a judicious reparametrization of the matter field. 
We subsequently put $h^{00}_\text{eff}$ identical to that of the original gauge by using the transformation (\ref{fieldrepara}) with $\gamma=\frac{2}{3}h^{00}$ 
\begin{align}
\langle h^{00}_\text{eff}h^{00}_\text{eff}\rangle_\text{de Donder}=\langle h^{00}h^{00}\rangle_0. 
\end{align}
Recall that the correlators in the original gauge are singled out by requiring Lorentz invariance of matter dynamics at the sub-horizon scale. 
We thus conclude that the scaling relation in (\ref{Scaling}) also holds in de Donder gauge once we impose effective Lorentz symmetry at sub-horizon scale. 

\section{Conclusion}
\setcounter{equation}{0}

We construct effective Lagrangian in de Sitter spacetime which summarizes IR logarithmic effects on local matter field dynamics. 
We integrate super-horizon mode of the metric fluctuations. 
We require Lorentz symmetry at sub-horizon scale. 
This strategy utilizes rescaling ambiguity of the matter fields with respect the conformal mode. 

The Lorentz symmetry at sub-horizon scale is a natural consequence of general covariance. 
We argue that IR logarithmic effects cannot spoil this fundamental symmetry. 
From microscopic physics point of view, general covariance is required from Lorentz symmetry and unitarity for quantum gravity. 
We thus argue that our requirement must be equivalent to the unitarity requirement. 

In de Sitter space, Lorentz symmetry is violated at horizon scale. 
However Lorentz symmetry holds at the sub-horizon scale in accordance with general covariance. 

The important consequence of our strategy is that we obtain unique physical predictions. 
In particular, we find that the couplings of $\phi^4$, Yukawa and gauge interactions are decaying with time. 
The screenings of these couplings can be expressed as logarithmic dependences of the scale factor and then they are slowly-progressing with cosmic evolution. 
Furthermore the relative evolution speed of the couplings are gauge independent once we impose Lorentz invariance of matter dynamics at the sub-horizon scale. 

The gauge change of metric propagator is equivalent to the coordinate change of the observer. 
However the observer can choose a new coordinate in which Lorentz symmetry holds at sub-horizon scale. 
This is realized by a reparametrization of matter fields. 
This procedure cancels the gauge change of the metric propagator as the matter-metric interaction is invariant under the reparametrization of both metric and matter fields. 
It is clear that a local quantity in quantum gravity depends on the observer. 
We need to specify the coordinate system of the observer to obtain gauge independent quantity. 
The requirement of Lorentz symmetry specifies such a system. 

\section*{Acknowledgment}

This work is supported by Grant-in-Aid for Scientific Research (B) No. 26287044 and (C) No. 16K05336. 
We thank S. Iso, H. Kawai and T. Tanaka for discussions. 

\appendix
\section{Propagators in a generalized gauge}\label{A:A}
\setcounter{equation}{0}

In the generalized gauge (\ref{generalGF}), the deformation from the original gauge fixing term $\delta=0$ is 
\begin{align}
\delta \mathcal{L}_\text{GF} \simeq \frac{1}{2}a^2\delta\big[&\ \eta^{\mu\nu}\partial_\mu h^{00}\partial_\nu h^{00}-3\partial_0 h^{00}\partial_0 h^{00}-\frac{5}{9}\partial_i h^{00}\partial_i h^{00} \notag\\
&-\frac{4}{3}\partial_i h^{00}\partial_k \tilde{h}^{ki}+\partial_k\tilde{h}^k_{\ i}\partial_l\tilde{h}^{li}\big]. 
\label{beta1}\end{align}
Here we have set $D=4$ and neglected massless conformally coupled modes. 
In addition, we have ignored ghost fields since they do not couple to matter fields. 
To investigate the gauge dependence, we evaluate the correction to the gravitational propagator
from the additional term (\ref{beta1}). 
We may focus on the $\delta$ dependent part: 
\begin{align}
\mathcal{L}_\text{eff}\simeq a^2\big[&-\frac{2}{3}(1-3\delta )\partial_0h^{00}\partial_0h^{00}+\frac{2}{3}(1-\frac{1}{3}\delta )\partial_ih^{00}\partial_ih^{00} \notag\\
&+\frac{1}{2}\partial_0S\partial_0S-\frac{1}{2}(1+\frac{4}{3}\delta )\partial_iS\partial_iS+\frac{4}{3\sqrt{3}}\delta \partial_ih^{00}\partial_iS \notag\\
&+\frac{1}{2}\partial_0V^i\partial_0V^i-\frac{1}{2}(1+\delta)\partial_jV^i\partial_jV^i\big], 
\end{align}
where $S$ and $V^i$ denote scalar and vector mode of $\tilde{h}^{ij}$ respectively \cite{Kitamoto2013}.
It can be canonically renormalized as
\begin{align}
\mathcal{L}_\text{eff}\simeq a^2\big[&-\frac{1}{2}\partial_0H\partial_0H+\frac{1}{2}A\partial_iH\partial_iH 
+\frac{1}{2}\partial_0S\partial_0S-\frac{1}{2}B\partial_iS\partial_iS+D \partial_iH\partial_iS \notag\\
&+\frac{1}{2}\partial_0V^i\partial_0V^i-\frac{1}{2}(1+\delta)\partial_jV^i\partial_jV^i\big], 
\end{align}
where 
\begin{align}
A=\frac{1-\frac{1}{3}\delta}{1-3\delta},\hspace{1em}B=1+\frac{4}{3}\delta,\hspace{1em}D=\frac{2}{3}\frac{\delta}{\sqrt{1-3\delta}},\hspace{1em}H=\frac{2}{\sqrt{3}}\sqrt{1-3\delta}h^{00}. 
\end{align}
We may rotate $H\rightarrow iH$ and find the frequency matrices in $(H,S)$ subspace to $O(\delta^2)$ as 
\begin{align}
\hat{\omega}^2=
\begin{pmatrix}1+\frac{8}{3}\delta+8\delta^2 & -\frac{2}{3}i\delta \\
-\frac{2}{3}i\delta & 1+\frac{4}{3}\delta \end{pmatrix}. 
\end{align}
In order to find the two-point functions of $(H,S)$, we consider
\begin{align}
\hat{\omega}^{-3} \simeq
\begin{pmatrix} 1-4\delta+\frac{1}{2}\delta^2 & i\delta\\
i\delta & 1-2\delta +\frac{5}{2}\delta^2 \end{pmatrix}. 
\end{align}
In this way, we find 
\begin{align}
\langle h^{00}h^{00}\rangle &= (1-\delta-\frac{5}{2}\delta^2)\times -\frac{3}{4}\langle \varphi^2\rangle, \notag\\
\langle\tilde{h}^{ij}\tilde{h}^{kl}\rangle&=(1-\delta +\frac{5}{4}\delta^2)\times(\delta^{ik}\delta^{jl}+\delta^{il}\delta^{jk}-\frac{2}{3}\delta^{ij}\delta^{kl})\langle\varphi^2\rangle. 
\end{align}

\section{Non-covariant field redefinition}\label{A:B}
\setcounter{equation}{0}

After the field redefinition: 
\begin{align}
\phi \to e^{\gamma\tau\partial_0}\phi 
=(1+\gamma\tau\partial_0+\frac{1}{2}\gamma^2\tau\partial_0+\frac{1}{2}\gamma^2\tau^2\partial_0^2)\phi, 
\label{r-scalar}\end{align}
and partial integrations, the action for the scalar field is rewritten as 
\begin{align}
S&=\int\sqrt{-g}d^4x\ \big[-(1+\gamma+\frac{1}{2}\gamma^2)g^{\mu\nu}\partial_\mu\phi\partial_\nu\phi^*-(2\gamma+4\gamma^2)g^{00}\partial_0\phi\partial_0\phi^* \notag\\
&\hspace{7em}-(\frac{1}{6}R+m^2)(1+3\gamma+\frac{9}{2}\gamma^2)\phi\phi^*-\frac{\lambda_4}{2}(1+3\gamma+\frac{9}{2}\gamma^2)(\phi\phi^*)^2\big] \notag\\
&=\int\sqrt{-g}d^4x\ \big[-e^{3\gamma}g^{00}\partial_0\phi\partial_0\phi^*-e^\gamma g^{ij}\partial_i\phi\partial_j\phi^* \notag\\
&\hspace{7em}-(\frac{1}{6}R+m^2)e^{3\gamma}\phi\phi^*-\frac{\lambda_4}{2}e^{3\gamma}(\phi\phi^*)^2\big]. 
\label{r-action1}\end{align}
We kept relevant terms up to the one-loop level and neglected differentiated gravitational fluctuations because they do not lead to IR effects.  

From (\ref{r-action1}), the non-covariant field redefinition (\ref{r-scalar}) can be identified as the modification of the way the gravity couples to the scalar field: 
\begin{align}
g_{00}\to g_{00},\ g_{ij}\to e^{2\gamma}g_{ij}\ \Leftrightarrow\ h^{00}\to h^{00}+\frac{3}{2}\gamma,\ \omega \to \omega+\frac{3}{4}\gamma.  
\label{r-interaction}\end{align}
Here we neglected the off-diagonal part of the metric $g_{0i}$ because its background value is zero and its fluctuation $h_{0i}$ does not lead to IR effects. 

Also redefining the other fields: 
\begin{align}
\psi\to e^{\gamma\tau\partial_0}\psi,\hspace{1em}A_0^a\to e^\gamma e^{\gamma\tau\partial_0}A_0^a,\hspace{1em}A_i^a\to e^{\gamma\tau\partial_0}A_i^a, 
\label{r-others}\end{align}
we can show that the interaction between the gravity and each matter field is modified according to (\ref{r-interaction}). 
Specifically, (\ref{r-scalar}) and (\ref{r-others}) modify the action as 
\begin{align}
S=\int\sqrt{-g}d^4x\ \big[&-e^{3\gamma}g^{00}D_0\phi(D_0\phi)^*-e^\gamma g^{ij}D_i\phi(D_i\phi)^* \notag\\
&-(\frac{1}{6}R+m^2)e^{3\gamma}\phi\phi^*-\frac{\lambda_4}{2}e^{3\gamma}(\phi\phi^*)^2 \notag\\
&+ie^{3\gamma}\bar{\psi}e^0_{\ 0}\gamma^0(\partial_0-iA_0^aT^a)\psi+ie^{2\gamma}\bar{\psi}e^i_{\ j}\gamma^j(\partial_i-iA_i^aT^a)\psi \notag\\
&-ie^{3\gamma}\bar{\psi}e^i_{\ j}\gamma^j\Sigma^{0k}\omega_{i0k}\psi -m_fe^{3\gamma}\bar{\psi}\psi-\lambda_Ye^{3\gamma}\phi\bar{\psi}\psi+\text{(h.c.)} \notag\\
&-\frac{1}{2g^2}e^\gamma g^{00}g^{ij}F_{0i}^aF_{0i}^a-\frac{1}{4g^2}e^{-\gamma}g^{ik}g^{jl}F_{ij}^aF_{kl}^a\big], 
\end{align}
where $\Sigma^{bc}=\frac{1}{4}[\gamma^b,\gamma^c]$ and $\omega_{\mu bc}$ is the spin connection: 
\begin{align}
\omega_{\mu bc}\simeq Ha(e_{\mu b}e^0_{\ c}-e_{\mu c}e^0_{\ b}). 
\end{align}
Only $\omega_{i0k}=-\omega_{ik0}$ is nonzero in our approximation.

\end{document}